# Hot Roller Embossing for the Creation of Microfluidic Devices

S. H. NG, Z. F. WANG

Singapore Institute of Manufacturing Technology

71 Nanyang Drive

Singapore 638075

*Abstract*-We report on the hot roller embossing of polymer sheets for the creation of microfluidic structures. Measurements conducted on 100 μm features showed that the lateral dimensions could be replicated to within 2% tolerance, while over 85% of mould depth was embossed. Feature sizes down to 50 μm and feature depths up to 30 μm had been achieved. At lower temperatures, asymmetric pile up of polymer material outside embossed regions was observed with higher pile up occurring on the trailing side of the embossed regions.

## I. INTRODUCTION

Low-cost, disposable microfluidic chips have application in point of care (POC) and mass screening operations where large quantities are required. The "use and dispose" strategy is similar in other medical devices such as the plastic syringes and hypodermic needles so that contamination due to reuse and inadequate cleaning or sterilization is eliminated. Polymer has always been the material of choice when applications require mass quantities at low cost, due to its low material cost and the associated mass production processes such as injection moulding. Traditional polymers face some drawbacks in terms of maximum operating temperature, gas permeability, structural strength, optical properties and chemical resistance when compared to silicon and glass. Recent advancement in polymer chemistry has produced some high-performance polymers such as polyetheretherketone (PEEK) and cyclic olefin copolymer (CoC), that can withstand higher temperatures and hasher chemical environment.

Polymer replication techniques such as hot embossing, injection moulding and soft lithography have traditionally been popular for the manufacturing of microfluidic chips. All these are mould based methods for creating open microchannels and cavities on polymer surfaces. Subsequent bonding with another piece of polymer will form the embedded microchannels and cavities. Emerging technologies include roll replication techniques such as UV (ultraviolet) roller embossing and hot roller embossing. Roller replication techniques have their attractiveness in mass production and compatibility with other reel-to-reel processes such as gravure or flexo printing for the creation of conductive elements, and lamination for sealing and encapsulation. UV roller embossing uses UV light to cure a thin film of photopolymer while hot roller embossing relies on heat and pressure to form the polymer. While UV roller embossing has the advantage of lower pressure and no heating requirement, there are limitations in terms of film thickness, chemical resistance and optical properties, due to the nature of the process and the properties of the UV curable resin. Hot roller embossing has its issues as well in terms of higher operating temperature, higher contact pressure, speed and other issues. In this paper, some initial studies and considerations of the hot roller embossing will be reported.

There is some related research on hot roller embossing in the literature. Tan et al. [1] performed roller nanoimprint lithography on thin coatings of photoresist that are hundreds of nanometers in thickness. They reported the advantage of using less force over a large substrate. However, their sample size was limited to a couple of centimeters and their setup could not perform continuous operation. Schift et al. [2] reported on the surface structuring of textile fibres using roll embossing. Sub-micron periodic structures were transferred from a thin metal shim (the mould) onto the surface of polyester fibres that were 180 μm in diameter. While roller embossing can achieve very high production rates (in excess of 1 m/s for plastic foil of up to 2 m width), the replication of microstructures with deep relief (>1 μm) is not yet an established mass-production process [3]. The hot embossing of polymer foil, or even metallised polymer foil, is well established and widely used for mass-producing foils for markets ranging from wrapping and packaging to diffractive optical elements, but with feature relief depth of only up to 1 μm [4]. The fidelity of replicated nano- or microstructures is very dependent upon the aspect ratio (feature height to width). A general rule of thumb is that an aspect ratio of 1:1 can be replicated easily, 5:1 with care and 10:1 only with great difficulty. The issue with aspect ratio arises at both the moulding and demoulding steps of the process.

## II. HOT ROLLER EMBOSSING

While previous works focused on the nanometer or sub-micron scale, we work on the micrometre scale typical of microfluidics structures. In our approach (see Fig. 1), a thermoplastic sheet is passed between two rollers. The top roller is made of steel with a nickel film mould mounted while the bottom roller is a rubber support roller. Heat is supplied to the embossing interface through the mould and a clamping pressure is applied between the two rollers. As the substrate passes between the rollers, the mould features are embossed into it. Preheating of the substrate can also be conducted before it is fed into the rollers. The process is different from hot embossing where the polymer substrate is given ample time to heat up and cooling is carried out with the force still applied. Table 1 lists some differences between the conventional hot embossing [5] that uses a flat, rigid mould and hot roller embossing. Typically, conventional hot embossing can give a better replication of the mould because the polymer can be demoulded at a temperature below its glass transition temperature, minimizing any polymer flow after the mould is removed. The trade off is a longer process time as compared to hot roller embossing.

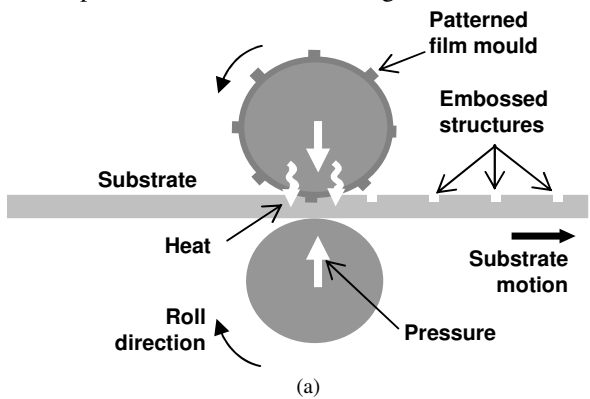

(a)

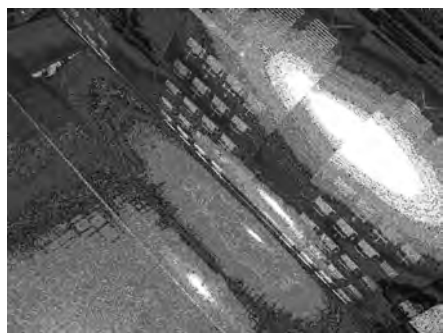

(b)

Fig. 1. Hot roller embossing process.

TABLE I
Some differences between conventional hot embossing that uses a flat, rigid mould and hot roller embossing.

|  | Conventional hot embossing | Hot roller embossing |
|---|---|---|
| Mould | Flat and rigid mould (typically electroformed nickel a few millimeters in thickness) | Typically a metal shim (tens to hundreds of microns thick) mounted on a rigid cylinder |
| Process | Piece by piece operation | Continuous |
| Temperature | Mould undergoes a temperature cycle of heating and cooling | Mould do not undergo temperature cycle |
| Load | Cyclical force over an area | Constant force operation, line loading |
| Demoulding | Mould and workpiece separation is in the direction perpendicular to the surface of the mould | Mould and workpiece separation similar to the peeling mode |

### III. EXPERIMENTAL

A 50 μm thick nickel mould (see Fig. 1(b)) with raised microstructures fabricated by electroplating process was wrapped around the top stainless steel roller that could be internally heated up to 177 °C. The fabrication of the film mould poses a challenge especially when the area is large. Figure 2 shows scanning electron micrographs of the features on the mould. The features consisted of line arrays with different line width and pattern density, as well as microchannel designs (not shown). The smallest feature size was 50 μm. The surface roughness of the mould base (see Fig. 3(a)) was ~ 0.1 μm. For better optical clarity, the surface roughness would have to be reduced further by optimizing the parameters or changing the scheme of the electroplating. Figure 3(b) shows a typical profile of mould feature. A thermocouple connected to a temperature controller was embedded into the roller. The machine had a web width of 450 mm and was capable of achieving a rolling speed of up to 0.1 m/s. Pneumatic pressure up to 6 bars (applied at two pneumatic pistons) provided the nip force between the rollers. Polymethyl methacrylate (PMMA) sheets with 1.5 mm thickness were used in the experiments. The PMMA had a glass transition temperature of ~ 105 °C measured from dynamic mechanical analysis.

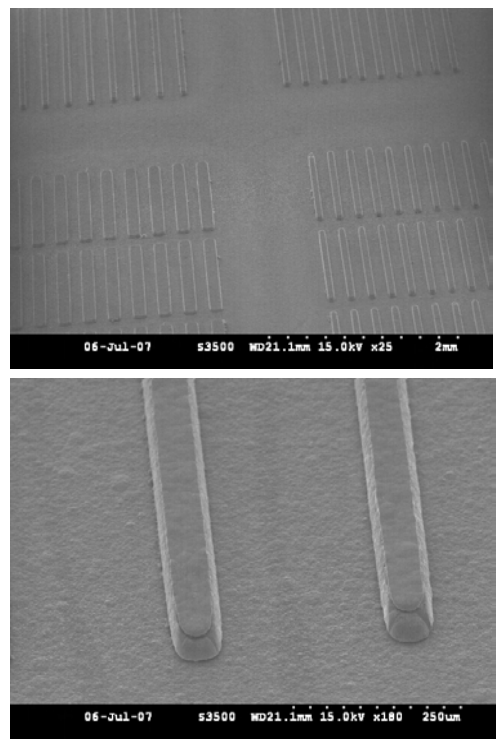

Fig. 2. Scanning electron micrographs of mould features.

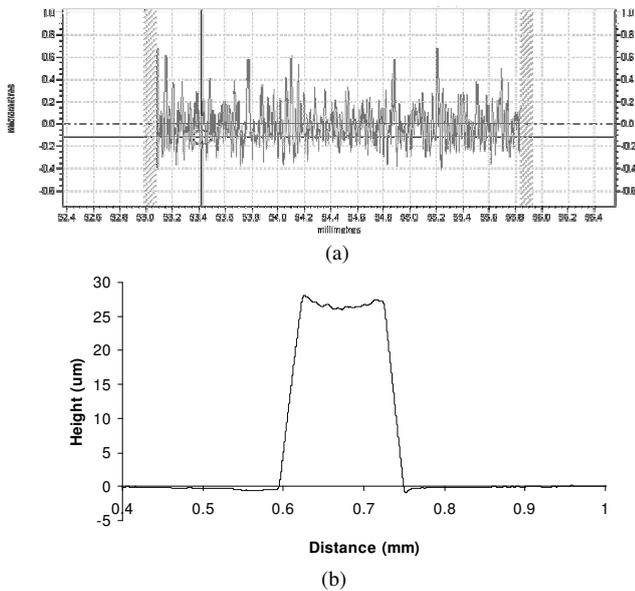

(a)

(b)

Fig. 3. Nickel electroplated mould features: (a) Surface roughness of mould base, (b) Two-dimensional profile of a feature.

## IV. RESULTS AND DISCUSSION

The primary parameters of the process include roller temperature, loading pressure and rolling speed. As seen in Fig. 4 and 5, the embossed depths increased when pressure is increased or when the roller speed is reduced. Figure 4(b) shows the actual nip force exerted between the rollers calibrated against the pneumatic pressure at the two pistons. In the graphs, the data points represent averages over 6 readings and the error bars are one standard deviation from the mean. The date was collected from a 100 μm line array using a stylus profilometer. The mould contained positive relief features occupying a small percentage of the mould base area. Hence, initial contact between the mould and the polymer happened at the top surface of the mould features. The mechanism is then similar to indentation where a higher contact pressure will result in a larger depth of penetration. The roller speed dictates the amount of time for transient heating of the surface of the polymer that contacts the mould. A slower speed will also allow a longer time for viscoplastic flow of the polymer material. As seen in Fig. 5(b), there is an optimal roller temperature (around 140 °C) in regards to the embossed depth. Below 110 °C, the embossed depth is less than 1 μm. Although the glass transition temperature of the PMMA is 105 °C, it is above 120 °C that the depth starts to increase rapidly, peaking at ~ 140 °C. This is related to the pressure applied and roller speed that determines the heating time of the polymer. An interesting phenomenon is seen above 140 °C where the depth starts to decrease. This could be a result of the partial reflow of the polymer material that has been heated up to a more fluidic state, after the mould separates from it. The phenomenon is not seen in rigid mould hot embossing where the demoulding is carried out at temperatures below the glass transition of the polymer, greatly reducing polymer movement after mould separation.

Figure 6(a) shows the profilometry scans of a 100 μm feature embossed at different temperatures from 80 °C to 130 °C. The depth of embossing increases with temperature. Larger increases in embossing depths were observed at temperature more than 110 °C. At the same time, pile up of material at the leading and trailing edges of the feature was observed (see Fig. 6(b)). There is a directional effect where the pile up is much higher at the trailing edge than at the leading edge. Both the pile up heights at the leading and trailing edges increase with temperature. At 130 °C, the leading edge pile up was ~ 4 μm while it was ~ 35 μm at the trailing edge. The pile up could be due to the shear forces acting at the mould and polymer interface as a result of the configuration of the roller embossing setup. The rubber support roller was driven by the motor. The embossing roller was in turn driven by the support roller when the two rollers came into contact. When the polymer sheet was fed between the two rollers, the embossing roller was driven by the polymer sheet. Hence, a shear force was acting on the embossing roller by the polymer sheet in the direction of the rolling motion. At the micro level, mould features that had indented into the polymer material experienced a horizontal force in the direction of the roller motion. The compressive stresses at the trailing face of the feature could have resulted in the larger built up of material, as compared to the pile up at the leading edge. With the increase in temperature, more pile up of polymer at this trailing edge occurred as a result of the decreased resistance to flow of the polymer material.

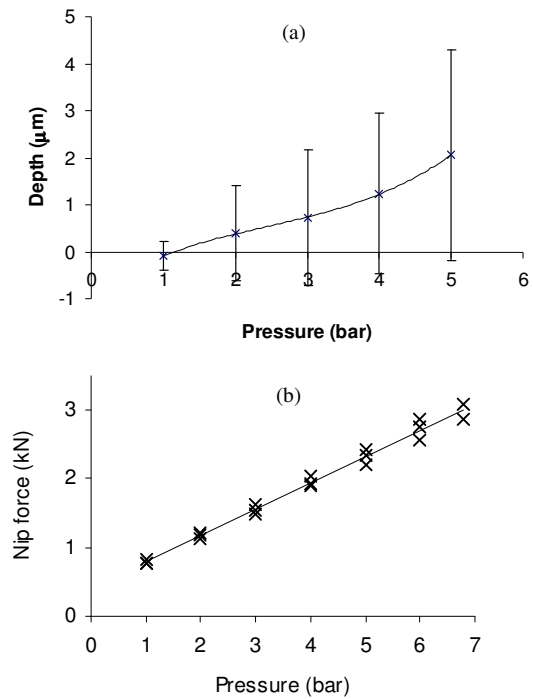

Fig. 4. Effect of load: (a) Embossing depth versus pneumatic pressure applied in the pistons, (b) Calibration curve of nip force versus pneumatic pressure.

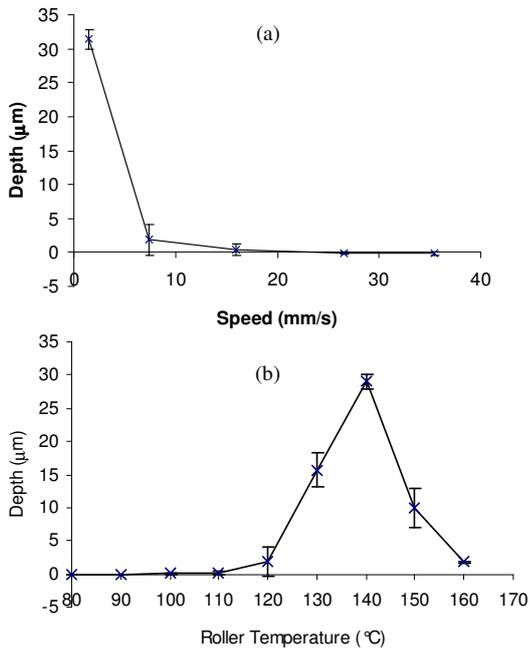

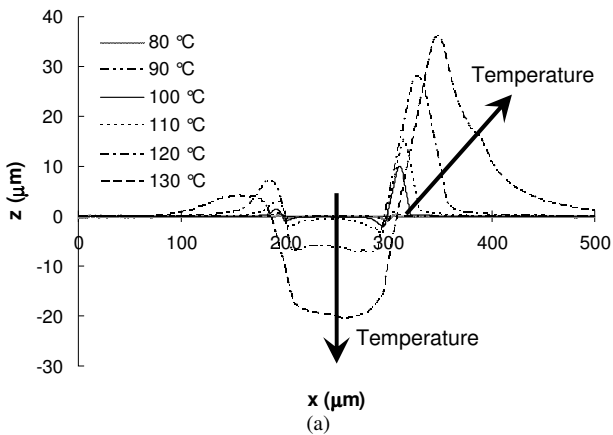

Measurements conducted on 100 µm features showed that the lateral dimensions can be replicated to within 2% tolerance, while over 85% of mould depth is embossed. Figure 7 shows 100 µm width raised lines on the mould in comparison to the corresponding embossed PMMA microchannels. Feature sizes down to 50 µm and feature depths up to 30 µm have also been demonstrated. Figure 8(a) shows a three-dimensional surface profile of an embossed design of a microchannel for capillary electrophoresis. The microchannel width is 100 µm and the average depth of embossing is ~ 20 µm. Some pile up can still be seen especially at the edges of the microchannel, as well as warpage of the chip itself. An array of the design was also adhesively bonded to another polymer to create embedded microchannels. Holes were drilled for fluid flow in and out of each feature. Figure 8(b) shows the array filled with a fluorescence dye in a simple flow test.

Fig. 5. Effects of (a) rolling speed and (b) roller temperature on embossed depth.

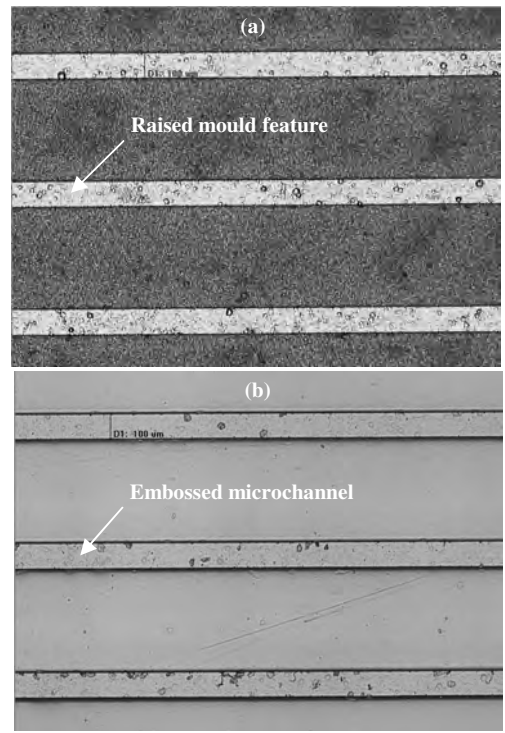

Fig. 7. (a) Mould features, (b) Embossed microchannels.

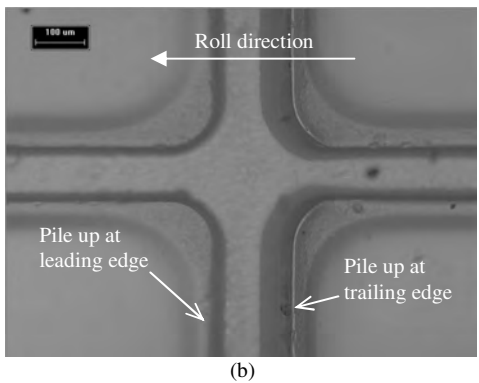

Fig. 6. Material pile up outside embossed regions: (a) Two-dimensional profiles of 100 µm features embossed at different temperatures showing the asymmetrical pile up of material, (b) Image showing feature embossed at 130 °C.

 

Thiele, and S. Westenhofer, "Replication technology for optical Microsystems," *Optics and Lasers in Engineering*, vol. 43, pp. 373, 2005.

[5]    S.H. Ng, Z.F. Wang, R.T. Tjeung, and N.F. de Rooij, "Development of a multi-layer microelectrofluidic platform," *Journal of Microsystem Technologies*, vol. 13, pp. 1509, 2007.

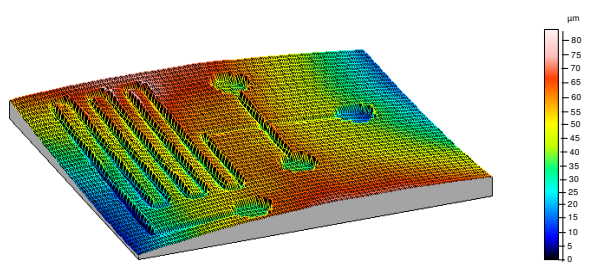

(a)

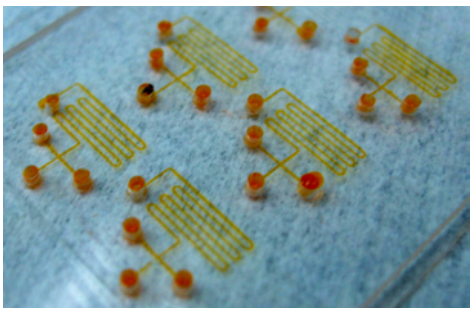

(b)

Fig. 8. Capillary electrophoresis separator fabricated by hot roller embossing: (a) Three-dimensional surface profile of roller embossed structure, (b) Bonded, embedded microchannels filled with fluorescence dye.

## V.    CONCLUSIONS

The results demonstrated hot roller embossing of PMMA sheets with embossed feature depths that were > 1 µm deep. This was possible with a compromise in the throughput. Embossed depths of more than 30 µm with 100 µm features were achieved at a speed of 1.5 mm/s. The embossed depths were within range of typical microchannel dimensions in microfluidic devices. The process could have the potential for mass production of polymer based microfluidic devices. In this research, feature sizes down to 50 µm and embossed depths of up to 30 µm had been demonstrated. An array of capillary electrophoresis separators were also created from the embossed structures. Flow tests were successfully conducted on this feature array.

## ACKNOWLEDGMENT

This research is funded by the Agency for Science, Technology and Research (A*STAR), Singapore.

## REFERENCES

[1]    H. Tan, A. Gilbertson, and S.Y. Chou, "Roller nanoimprint lithography," *Journal of Vacuum Science Technology B*, vol. 16, pp. 3926, 1998.

[2]    H. Schift, M. Halbeisen, U. Schutz, B. Delahoche, K. Vogelsang, and J. Gobrecht, "Surface structuring of textile fibers using roll embossing," *Microelectronic Engineering*, vol. 83, pp. 855, 2006.

[3]    H.P. Herzig, *Micro-optics: Elements, systems and applications*, London: Taylor and Francis, 1997.

[4]    M.T. Gale, C. Gimkiewicz, S. Obi, M. Schnieper, J. Sochtig, H.